\documentstyle [12pt]{article}
\begin{document}

\thispagestyle{empty}

\hphantom{x}

\vspace{1.5in}

\setlength{\unitlength}{1in}
\begin{center}
{\Large{\bf Momentum Distribution for Bosons with Positive Scattering
Length in a Trap}} \end{center}
\baselineskip 0.2in
\begin{center}
\ \\
\ \\
{\bf T.T. Chou}$^1$, {\bf Chen Ning Yang}$^2$ {\bf and L.H. Yu}$^3$\\ \ \\
$^1${\small\em{Department of Physics, University of Georgia, Athens,
Georgia 30602}} \\
$^2${\small\em{Institute for Theoretical Physics, State University of New
York}}, \\
{\small\em {Stony Brook, New York 11794}} \\

$^3${\small\em{National Synchrotron Light Source, Brookhaven National
Laboratory, Upton}}, \\
{\small\em {New York 11973}}\\
\ \\
\ \\
\end{center}

\begin{abstract}
\normalsize
The coordinate-momentum double distribution function $\rho  ({\bf r}, {\bf
p})  d^{3}rd^{3}p$
is calculated in the local density approximation for bosons with positive
scattering length $a$ in
a trap. The calculation is valid to the first order of $a$. To clarify the
meaning of the result, it is
compared for a special case with the double distribution function
$\rho_{w}d^{3} rd^{3}p$ of
Wigner.
\end{abstract}

\newpage
\baselineskip 0.30in
Using the local density approximation (LDA) [1,2], which is a
straightforward adaptation
of the Thomas-Fermi method, the density distribution $\rho({\bf r})d^{3}r$
in coordinate space
for BEC for $a > 0$ in a trap has been obtained. We want to calculate in
this paper the
coordinate-momentum distribution $\rho({\bf r}, {\bf p}) d^{3}rd^{3}p$ in
the same
approximation. We follow the notation of Ref. [2] throughout. In
particular, the fugacity $z$ of
the system is \begin{equation} z = {\rm exp}[\mu /kT]
\end{equation}
where $\mu$ is the chemical potential. We introduce a {\it local fugacity}
$\zeta({\bf r})$
defined as
\begin{equation}
\zeta = z \ {\rm exp}[-{\beta}V({\bf r})]. \end{equation}

\noindent {\bf 1. The Gaseous Phase}

By the gaseous phase we include both the system before BEC sets in, and the
system at high
densities for the cells outside of $r_{0}$ [2], i.e., outside of the region
where BEC takes place. We
consider such a cell of volume $V$ in which the local fugacity is $\zeta$.
Using the method of
Ref.[3] we write the grand partition function in the cell as
\begin{equation}
{\cal Q} = \sum_N \zeta^N Tr[{\rm exp}(-\beta H_0-\beta H')]. \end{equation}
The average occupation number $\ll n_k \gg$ of the state with momentum
$\hbar {\bf k}$ can be
computed from
\begin{equation}
{\cal Q} \ll n_k \gg = \sum_N \zeta^N Tr[(a^{\dagger}_{k} a_k){\rm exp}
(-\beta H_0-\beta H')].
\end{equation}
We shall drop all terms beyond the first order of $H'$. Since $H_0$
commutes with $a^{\dagger}_{k} a_k$, we find \begin{equation}
{\cal Q} = \sum_N \zeta^N Tr[{\rm exp}(-\beta H_0) (1-\beta H')] = {\cal
Q}_0 + {\cal Q}_1
\end{equation}
and
\begin{equation}
{\cal Q} \ll n_k \gg = \sum_N \zeta^N Tr[{\rm exp}(-\beta H_0)
(a^{\dagger}_{k} a_k)(1-\beta H')] = {\cal A}_{0} + {\cal A}_{1}
\end{equation}
where
\begin{equation}
{\cal Q}_0 = \sum_N \zeta^N Tr{[{\rm exp}(-\beta H_0)] = \prod [1- \zeta
e^{-\beta\varepsilon}]}^{-1} \end{equation}
is the term in ${\cal Q}$ without the perturbation term. ${\cal Q}_1$ has
been evaluated in Ref.[3]. \begin{equation}
{\cal Q}_1 / {\cal Q}_0 = -\beta \frac{4\pi a \hbar^{2}}{mV} \left[\sum_{\alpha
\neq\beta} \bar{n}_{\alpha} \bar{n}_{\beta} + \sum_{\alpha} \frac{1}{2}
{\overline {n^2_{\alpha}}} - \sum_{\alpha} \frac{1}{2} \bar{n}_{\alpha}
\right]
\end{equation}
where the bar means average over the grand canonical ensemble ${\cal Q}_0$:
\begin{equation}
\bar{n}_{\alpha} = \frac{\zeta e^{-\beta \varepsilon_{\alpha}}} {1- \zeta
e^{-\beta \varepsilon_{\alpha}}} \ \ . \end{equation}
Similarly
\begin{equation}
{\cal A}_0 / {\cal Q}_0 = \bar{n}_k
\end{equation}
and
\begin{equation}
{\cal A}_1 / {\cal Q}_0 = -\beta \frac{4\pi a \hbar^{2}}{mV} \left< n_{k}
\left[\sum_{\alpha \neq\beta}{n}_{\alpha} {n}_{\beta} + \sum_{\alpha}
\frac{1}{2} {n}^{2}_{\alpha} - \sum_{\alpha} \frac{1}{2}{n}_{\alpha}
\right] \right>
\end{equation}
where the symbol $\langle \ \ \rangle$ means the same average as the bar.
The coefficient in (8) and (11), $-\beta 4 \pi a \hbar^{2} (mV)^{-1}$, is
equal to $-2a \lambda^{2} V^{-1}$. Now (11) can be rewritten as
\begin{equation}
{\cal A}_1 / {\cal Q}_0 = -2a \lambda^{2} V^{-1} \left< (n_{k} - \bar{n}_{k})
\left[\sum_{\alpha \neq\beta}{n}_{\alpha} {n}_{\beta} + \sum_{\alpha}
\frac{1}{2} {n}^{2}_{\alpha} - \sum_{\alpha} \frac{1}{2}{n}_{\alpha}
\right] \right> + \bar{n}_{k} {\cal Q}_{1} / {\cal Q}_{0}. \end{equation}
Notice that ${\cal Q}_{0}$ is a product distribution function according to
(7). Thus $\overline{n_{\alpha} n_{\beta}} = \bar{n}_{\alpha}
\bar{n}_{\beta}$ if $\alpha \neq \beta$. Using this and similar identities
we find that in the sum over $\alpha$ and $\beta$ in (12), the bracket
$\langle \ \ \rangle$ vanishes unless $k = \alpha$ or $k = \beta$. Thus
\begin{equation}
{\cal A}_1 / {\cal Q}_0 = -2a \lambda^{2} V^{-1} \left< \sum_{\beta \neq k}
2n_k n_{\beta}(n_k - \bar{n}_k) + \frac{1}{2} (n^{3}_{k} - \bar{n}_k
n^{2}_{k} - n^{2}_{k} + \bar{n}^{2}_{k}) \right> + \bar{n}_k {\cal Q}_{1} /
{\cal Q}_{0}. \end{equation}
Now $V^{-1} \left< \sum_{\beta \neq k} n_k n_{\beta} (n_k - \bar{n}_{k})
\right> \rightarrow \rho (\overline {n^{2}_{k}} - \bar{n}^{2}_{k})$ as $V
\rightarrow \infty$, yielding \begin{equation}
{\cal A}_1 / {\cal Q}_0 = -4a \lambda^{2} \rho (\overline{n^2_k} -
\bar{n}^{2}_{k}) + \bar{n}_{k} \ {\cal Q}_1 / {\cal Q}_0. \end{equation}
Adding this to (10) and dividing by $1 + {\cal Q}_1 / {\cal Q}_0$ we
obtain, to order $a$,
\begin{equation}
\ll n_{k} \gg = \bar{n}_{k} - 4a \lambda^{2} \rho (\overline{n^2_k} -
\bar{n}^{2}_{k}).
\end{equation}
The number of modes ${\bf k}$ in a cell of volume $V$ is ${(8
\pi^{3})}^{-1} V d^{3}k$. Thus the
combined coordinate-momentum distribution $\rho ({\bf r}, {\bf p})$ is given by
\begin{equation}
h^{3} \rho({\bf r}, {\bf p}) = \ll n_k \gg  = \frac{\zeta e^{-\beta
\varepsilon_{\alpha}}} {1- \zeta
e^{-\beta
\varepsilon_{\alpha}}} -4 \pi \lambda^{2} \rho(r) \frac{\zeta e^{-\beta
\varepsilon_{\alpha}}}
{{(1- \zeta e^{-\beta \varepsilon_{\alpha}}})^{2}} \end{equation} where
$\varepsilon_k =
\frac{\hbar^2 k^2}{2m}$ and $\zeta$ is given by (2). In (16) we have evaluated
$\overline{n^2_k}$ in a straightforward way from the product partition
function (7).

Integrating (16) over $d^3 p$ we should get the density $\rho(r)$ times
$h^3$. This can be done without much difficulty, yielding Eq. (3) of Ref.
[2].

\noindent {\bf 2. The Region with Condensate}

For high densities, BEC forms in some cells of the trap. In those cells
$\rho = \rho_0 + \rho_s > \rho_0$, where [4], \begin{equation}
\rho_0 = \lambda^{-3} g_{3/2} (1)
\end{equation}
and
\begin{equation}
V(r) + 4 \pi a \rho_s (r) \hbar^2 / m = V(r_0). \end{equation}
Here $\rho_s$ denotes superfluid density, i.e., density of particles with
${\bf p} = 0$.
An important parameter $\xi_5 = \rho_s / \rho$, a function of the location
of the cell, with value
between 0 and 1, describes {\em incomplete occupation} of the ground state,
and was studied in
detail in Ref. [5]. [Notice that $\xi_5$ and $\xi$ are totally different
quantities.] For cells without
BEC, $\xi_5 = 0$.

It was shown in Ref. [5] that the system in a cell with BEC has an energy
given by (5.16)
with a {\em phonon} spectrum (for $k \neq 0$) given by (5.18):
\begin{equation}
\hbar \omega_k = \frac{\hbar^2}{2m} {(k^4 + 2k^{2}_{0} k^2)}^{1/2}, \ \
k^2_0 = 8 \pi a \xi_5
\rho = 8 \pi a \rho_s. \end{equation}
Notice that for the gaseous phase, $\xi_5 = 0$ and the phonon spectrum is
quadratic for
small $k$.

The phonon creation operator $b^\dagger_k$ and the particle creation
operator $a^\dagger_k$
are related to each other through a Bogoliubov transformation [6]:
\begin{equation}
a_k = (b_k - \alpha_k b^{\dagger}_{-k}) {(1- \alpha^2_k)}^{-1/2} \end{equation}
where
\begin{equation}
\alpha_k = k^{-2}_0 (k^2 + k^2_0 - \sqrt{k^4 + 2k^2 k^2_0}). \end{equation}
For a state with $m_k$ phonons we can compute the average occupation number
$\ll n_k \gg$
of atoms in the state ${\bf p} = \hbar {\bf k}$ using (20) above. The
result is linear in $m_k$. Now
the average number of $m_k$ is given by Eqs.(5.27) and (5.31). Thus
\begin{equation}
\rho({\bf r}, {\bf p}) = h^{-3} [\alpha^2_k + (1 + \alpha^2_k) {(e^{\beta \hbar
\omega_k} - 1)}^{-1}] {(1 - \alpha{^2_k})}^{-1}, \ \ (k \neq 0), \ \
\end{equation}
where $\omega_k$ is given by (19), and $\alpha_k$ is given by (21).

For $k \gg k_0 = \sqrt{8 \pi a \rho_s}$, the phonon energy (19) can be
expanded in powers of
$a$ and (22) becomes \begin{equation}
\rho ({\bf r}, {\bf p}) = h^{-3} {(e^{\beta E_k} - 1)}^{-1}, \ \ (k \gg k_0),
\end{equation}
where
\begin{eqnarray*}
E_k = \frac{p^2}{2m} + \frac{\hbar^2}{2m} [8 \pi a \rho_s (r)]. \end{eqnarray*}
For other values of $ k > 0$, Eq.(22) gives the distribution. It is a
complicated function of $k$.
For $0 < k \ll k_0$, it reduces to
\begin{equation}
\rho({\bf r}, {\bf p}) \cong h^{-3} m \beta^{-1} p^{-2}, \ \ (0 < k \ll
k_0). \end{equation}
Notice that this differs by a factor of 2 from the corresponding
distribution when $a=0$.

\noindent {\bf 3. Wigner Double Distribution}

What is the meaning of the double distribution $\rho({\bf r}, {\bf p})$?
It, of course, should only
be used [2] for
$d^3 r > {(L_2)}^3$, and for $d^3 r d^3 p > h^3$. But does it have a clear
meaning in quantum
mechanics? We discuss this by examining Eq.(16) in the limit of $a=0$, for
the case of a
spherically symmetrical harmonic trap $V(r) = \frac{1}{2} m \omega^2 r^2$.
In such a case we
can compute exactly the matrix element of $\langle {\bf r}' | \frac{z
e^{-\beta H}} {1- z
e^{-\beta H}} | {\bf r} \rangle = \sum^{\infty}_{\ell = 1} \langle {\bf r}'
| z^{\ell} e^{-\beta \ell H}
| {\bf r} \rangle$, by using, e.g., the result of Ref.[9]. Using Wigner's
idea [10], we put ${\bf r}' =
{\bf R} - \frac{1}{2}$ $\boldmath{\eta}$, and ${\bf r} = {\bf R} + \frac{1}{2}$
$\boldmath{\eta}$ and  evaluate the
above, and then make a Fourier transform to the variable ${\bf P}$
conjugate to $\boldmath{\eta}$.

The resultant double distribution $a \;\ell a$ Wigner becomes
\begin{equation}
\rho_w({\bf R}, {\bf P}) = h^{-3} \sum^{\infty}_{\ell = 1} z^{\ell} ({\rm
sec} h \frac{\ell
\varepsilon}{2}){^3} \ {\rm exp} \left\{ - \frac{2 \beta}{\varepsilon}
({\rm tan}h \frac{\ell
\varepsilon}{2}) (\frac{P^2}{2m} + \frac{1}{2} m \omega^2 R^2) \right\}
\end{equation} where
$\varepsilon = \beta \hbar \omega$. In the limit that $\varepsilon
\rightarrow 0$, this is exactly
Eq.(16) for $a=0$, [noticing that the local fugacity $\zeta$ is given by
(2)] which is in agreement
with the discussion in Ref. [2] for the single distribution function $\rho(r)$.

The work of CNY is supported in part by an NSF Grant PHY-9309888. That of
LHY is performed under the auspices of US DOE.

\newpage
\begin{center}
{\bf REFERENCES}
\end{center}
\begin{enumerate}
\item J. Oliva, Phys. Rev. {\bf B39}, 4197 (1989). \item T. T. Chou, Chen
Ning Yang and L. H. Yu, to appear in
Phys. Rev. A.
\item K Huang, C. N. Yang and J. M. Luttinger, Phys. Rev.
{\bf 105}, 776 (1957).
\item In Ref. [2] we only discussed cases where $V(r)$ is spherically
symmetrical. But it is obvious that the discussion can be extended to
nonspherically symmetrical
cases. \item T. D. Lee and C. N. Yang, Phys. Rev. {\bf112}, 1419 (1958).
To avoid confusion of notations between Refs. [2] and [5], we add a
subscript 5 to the
parameters $\xi$ and $\zeta$ of Ref. [5]. Furthermore we refer to equation
x in Ref. [5] as (5.x).
\item N. N. Bogoliubov, J. Phys. USSR, {\bf XI}, 23 (1947), first
introduced the mathematical trick
later known as the Bogoliubov transformation. But the concept of scattering
length was not used
and the physics of the paper was incorrect. In 1957 in Ref. [7] the physics
of the dilute hard
sphere boson system was reduced to a Hamiltonian problem solved in the appendix of that paper.
Later it was pointed out [8] that this Hamiltonian problem can be more
easily solved with the
Bogoliubov transformation. We follow here this later method.
\item T. D. Lee, K. Huang and C. N. Yang, Phys. Rev. {\bf 106},
1135 (1959).
\item K. Huang, {\em Statistical Mechanics} (John Wiley, New York 1963).
\item R. P. Feynman, {\em Statistical Mechanics: a Set of Lectures}
(Benjamin, New York, 1972).
\item E. Wigner, Phys. Rev. {\bf 40}, 749 (1932).

\end{enumerate}
\end{document}